\documentclass[runningheads]{llncs}
\usepackage[T1]{fontenc}
\usepackage{booktabs}
\usepackage{stmaryrd}
\usepackage{amssymb}
\usepackage{amsfonts}
\usepackage{placeins} 
\usepackage{cite}
\AtBeginDocument{%
  }

\usepackage{graphicx}
\usepackage{tcolorbox}
\usepackage{dialogue}
\usepackage[skip=5pt]{caption}
\usepackage[capitalise, noabbrev]{cleveref}
\usepackage{float}
\usepackage{xcolor} 

\begin{document}

\title{Dude, where's my utterance? Evaluating the effects of automatic segmentation and transcription on CPS detection
}

\author {Videep Venkatesha\orcidID{0009-0000-4635-3010} \and Mariah Bradford\orcidID{0009-0009-2162-3307} \and Nathaniel Blanchard\orcidID{0000-0002-2653-0873}}
\authorrunning{V. Venkatesha et al.}
\titlerunning{Evaluating CPS Detection with ASR and Segmentation}
%
\institute{Colorado State University, Fort Collins, CO 80523, USA \
\email{videep@rams.colostate.edu}
}

\maketitle
\begin{abstract}

Collaborative Problem-Solving (CPS) markers capture key aspects of effective teamwork, such as staying on task, avoiding interruptions, and generating constructive ideas. An AI system that reliably detects these markers could help teachers identify when a group is struggling or demonstrating productive collaboration. Such a system requires an automated pipeline composed of multiple components.
In this work, we evaluate how CPS detection is impacted by automating two critical components: transcription and speech segmentation. On the public Weights Task Dataset (WTD), we find CPS detection performance with automated transcription and segmentation methods is comparable to human-segmented and manually transcribed data; however, we find the automated segmentation methods reduces the number of utterances by 26.5\%, impacting the the granularity of the data. We discuss the implications for developing AI-driven tools that support collaborative learning in classrooms.

\end{abstract}




%

\section{Introduction}

A widely used approach for assessing group work is through identifying Collaborative Problem-Solving (CPS) markers. CPS is a key 21st-century skill emphasized in frameworks like PISA 2015 \cite{oecd2017pisa,thinkkids,klieme2016assessing}. CPS markers categorize group interactions based on behavioral indicators. These markers, as outlined in  \cite{sun_towards_2020}, are divided into three primary facets: (1) Constructing Shared Knowledge, where students build common ground and exchange information; (2) Negotiation and Coordination, where they propose solutions, refine ideas, and resolve conflicts; and (3) Maintaining Team Function, which involves regulating interactions and ensuring productive collaboration. 

Previous studies have explored CPS in diverse contexts, including work-based learning \cite{jensen2023collaborative} \cite{oecd2017pisa, klieme2016assessing}. Additionally, \cite{Flor2016} introduced computational linguistic methods for automatically characterizing students’ CPS skills, demonstrating that automated approaches could serve as viable alternatives to human-driven methods. \cite{bradford_automatic_2023} further established that CPS markers could be detected using multimodal features in naturalistic, unconstrained group settings, providing foundational work for automated classroom monitoring. However, while their work demonstrated the viability of automated CPS detection broadly, it does not explore the impact and nuances of using a fully automated system.

\begin{figure}
  \centering
  \includegraphics[width=0.9\linewidth]{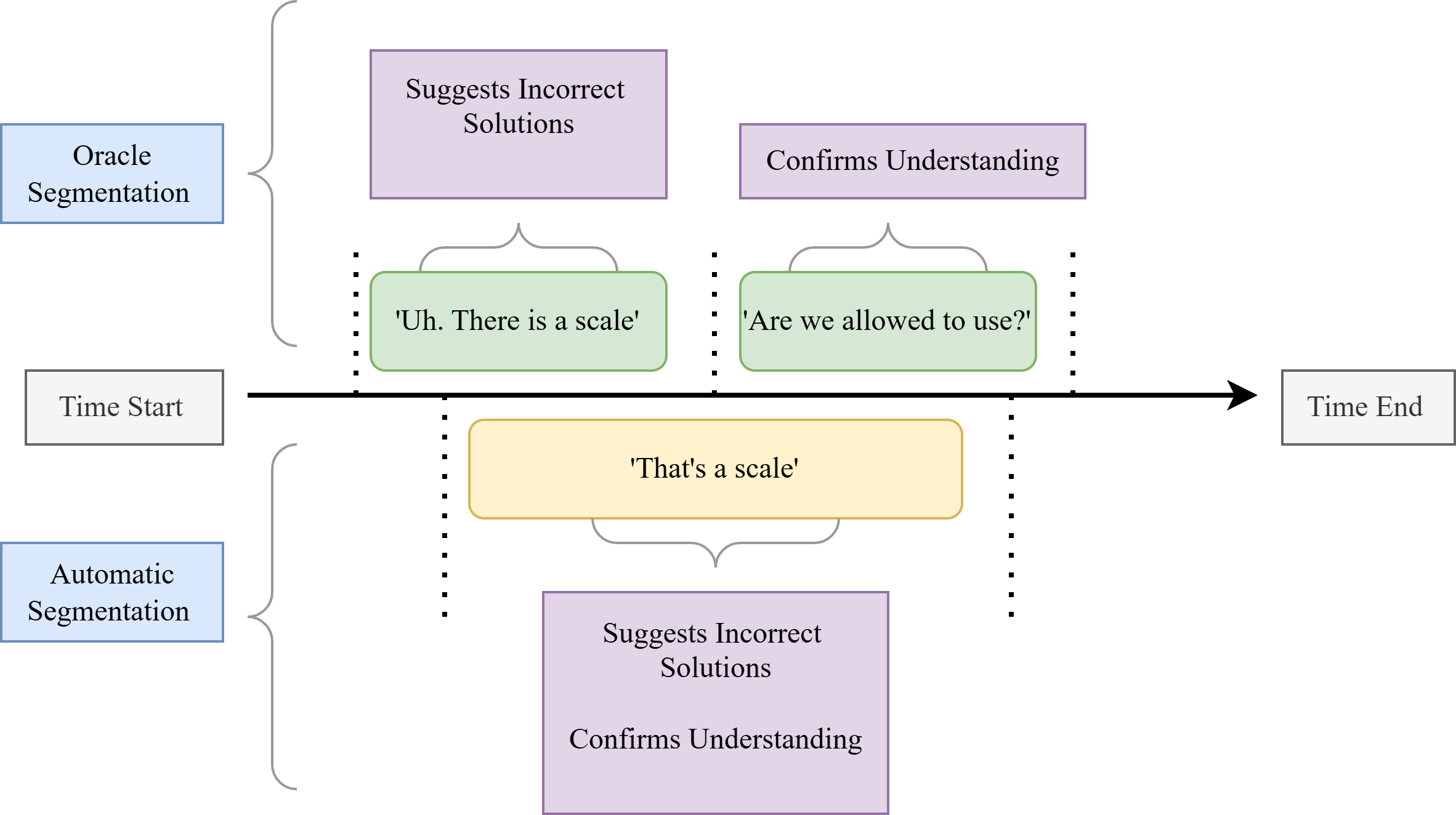} 
  \caption{Comparison of Oracle segmentation (green) and Google automatic segmentation (yellow) over the same time span. Oracle preserves distinct utterances, transcripts, and labels; Google merges them, reducing granularity.}
  \label{fig:segments}
\end{figure}

To achieve fully automated CPS detection, multiple interconnected components must be carefully considered, each contributing significantly to the reliability of the final system. Crucially, these components include segmentation methods and transcription quality, both of which substantially influence downstream classification performance. Figure \ref{fig:segments} illustrates this issue: over the same time window, the Oracle segmentation captures two separate utterances, each with distinct content and collaborative function, while the automated system merges them into one. This leads to both label overlap and transcript compression, which can obscure the speaker turns and degrade downstream CPS classification

Segmentation, in this context, refers to the division of continuous group audio into discrete speaker-aligned utterances. The specific impact of using automated segmentation and transcription on the accuracy and granularity of CPS detection remains under-explored in current literature. While Automatic speech recognition (ASR) systems have been extensively evaluated in various contexts \cite{gales2008hmm, yu2014asr, yoon2020asr, zechner2015asr, manuvinakurike2015asr, evanini2013asr}, few studies specifically assess ASR within classroom discourse \cite{bradford2022deep, blanchard2015study}. Cao et al. \cite{cao2023comparative} found that ASR errors severely impacted lexically sensitive tasks but had less effect on discourse-level classification. However, their study focused broadly on discourse modeling rather than explicitly on CPS, which requires a more nuanced, finer-grained analysis of collaborative interactions.

In this paper, we directly address the research gap by investigating the following research question: How does the use of automated transcription and segmentation affect the detection of CPS markers in small group dialogues? Our findings indicate that fully automated CPS detection pipelines, utilizing ASR-based transcription and segmentation, do not significantly degrade performance compared to human-segmented and manually transcribed data. However, this comes with the caveat that automated methods may not capture CPS behaviors at a granular level, potentially merging distinct utterances and reducing interpretability.

\section{Methodology}

\subsection{Dataset: The Weights Task Dataset}

We utilize the Weights Task Dataset (WTD) \cite{khebour2024text}, which consists of ten triads engaged in a structured problem-solving activity, where participants must determine the weights of a set of blocks using a balance scale. Each interaction is annotated for CPS indicators, following the framework outlined by \cite{sun_towards_2020}. This framework includes 19 distinct markers of collaborative behavior, with each of them being mapped to  constructing shared knowledge, negotiation/coordination, and maintaining team function. The dataset includes Gold-standard (Oracle) segmentation and transcriptions which were manually annotated to serve as the most accurate reference and Google Automatic Speech Recognition (ASR) transcriptions, which was automatically segmented using Google's Voice Activity Detector. 

\subsubsection{Feature Extraction}

We extracted both linguistic and acoustic modalities from the dataset. We extracted linguistic features using BERT-base-small and prosodic features using openSMILE. These were concatenated to form a compact multimodal representation of each utterance, following prior work \cite{openFeatures,devlin2018bert,eyben2010opensmile}.


\subsection{Modeling Approach}

We employ a supervised learning framework to classify Collaborative Problem Solving (CPS) facets. The classification task is framed as a multi-label classification problem, where each utterance can belong to one or more CPS facets: Constructing Shared Knowledge (Const), Negotiation and Coordination (Neg), and Maintaining Team Function (Maintain).

\subsubsection{CPS Indicator Labels}
Each utterance was assigned to a CPS facet if it contained any subcategory corresponding to that facet.  
For example, if an utterance contained both \textit{Suggests appropriate ideas} and \textit{Confirms understanding}, it would be labeled as Const = 1, while Neg = 0 and Maintain = 0, since these two facets fall under \textit{Constructing Shared Knowledge}. This ensured that all relevant utterances contributed to their respective high-level categories. The Google segments inherited the labels of all of the oracle segments that fell under the time span of the Google segment. 

\subsubsection{Classification Models}

We experimented with Random-Forest \cite{breiman2001random}, and AdaBoost \cite{freund1997decision} following the methodology in \cite{bradford_automatic_2023} to evaluate the effectiveness of different classification models. Deep learning models were not considered, as the size of the dataset was insufficient to support their effective training without overfitting.
Hyperparameter tuning was performed using \textit{Hyperopt}, optimizing for \textit{average AUROC scores} over 500 iterations \cite{bergstra2013hyperopt}. The search was conducted using the Oracle-Segmented and Oracle-Transcripts condition, and the best-performing model was then evaluated across all other conditions. The best model identified was a Random Forest classifier with criterion set to \texttt{entropy}, max features set to \texttt{None}, and number of estimators set to \texttt{148}.  We employ leave-one-group-out cross-validation to evaluate model performance. This is a common technique used to ensure robustness of the machine learning model \cite{efron1993bootstrap}. We assess model performance using AUROC (Area Under the Receiver Operating Characteristic Curve).  AUROC values range from 0 to 1, where 0.5 is random guessing and 1 is perfect  \cite{fawcett2006introduction}. Aditionally, we also calulate the average Precision which measures the proportion of predicted positive labels that are actually correct, and Recall which measures the proportion of actual positive labels that were correctly identified, of the 3 CPS facets, for all conditions. 
\section{Results}
\label{sec:results}

Table~\ref{tab:segmentation_transcription} presents AUROC scores for different combinations of segmentation and transcription methods. 
As expected, the highest AUROC is achieved with Oracle transcripts and segmentation (average AUROC = 0.744), while the fully automated condition (Google segmentation and transcription) yields the lowest performance (average AUROC = 0.679). Notably, using Google segmentatizon with Oracle transcripts nearly matches the Oracle–Oracle performance (0.740 vs. 0.744). Exploring each condition further provides deeper insight into the nature of these effects. Table~\ref{tab:precision_recall_segmentation_transcription} presents the average precision and recall across all CPS facets.

\begin{table*}[h]
\centering
\caption{AUROC scores by segmentation and transcription method. Highest values in bold.}
\begin{tabular}{lllllc}
\toprule
\textbf{Seg.} & \textbf{Tran.} & \textbf{Const($\pm$SD)} & \textbf{Neg($\pm$SD)} & \textbf{Maintain($\pm$SD)} & \textbf{Average} \\ \midrule
Oracle & Oracle & \textbf{0.758 (0.044)}  & \textbf{0.765 (0.045)} & \textbf{0.720 (0.045)} & \textbf{0.744} \\ 
Oracle & Google & 0.721 (0.029)  & 0.720 (0.049) & 0.653 (0.031) & 0.698 \\ 
Google & Oracle & 0.754 (0.066)  & 0.757 (0.047) & 0.699 (0.061) & 0.740 \\ 
Google & Google & 0.718 (0.079)  & 0.705 (0.071) & 0.616 (0.078) & 0.679 \\ 

\bottomrule
\end{tabular}
\label{tab:segmentation_transcription}
\end{table*}

We observe that automated transcription substantially degrades precision (from 0.704 to 0.528) and recall (from 0.297 to 0.246) when manual high-quality segmentation is used. This suggests that transcription quality plays a critical role in preserving classifier performance. Automated segmentation has a more nuanced impact: when using oracle transcripts, precision drops slightly (from 0.704 to 0.658), but recall improves (from 0.297 to 0.339), suggesting that merged utterances may help capture broader CPS behaviors at the cost of granularity.

\begin{table*}[t]
\centering
\caption{ Average precision and recall across CPS facets. Highest values in bold.}
\begin{tabular}{llcc}
\toprule
\textbf{Seg.} & \textbf{Tran.} & \textbf{Avg. Precision} & \textbf{Avg. Recall} \\ \midrule
Oracle & Oracle &  \textbf{0.704} & 0.297 \\ 
Oracle & Google &  0.528 & 0.246 \\ 
Google & Oracle &  0.658 & 0.339 \\ 
Google & Google &  0.601 & \textbf{0.342} \\ 
\bottomrule
\end{tabular}
\label{tab:precision_recall_segmentation_transcription}
\vspace{-10pt}
\end{table*}

Interestingly, the fully automated condition (Google segmentation and transcription) achieves the highest recall (0.342) despite lower precision (0.601), indicating potential utility for real-time applications that prioritize broader detection. The poorest-performing condition is Oracle segmentation with Google transcription—likely due to a mismatch between clean segment boundaries and noisy transcriptions.
\vspace{-10pt}
\section{Discussion}

A striking finding of this study is that the use of automated segmentation and ASR-based transcription does not seem to significantly degrade CPS classification performance compared to oracle transcriptions and human-segmented data. This result is particularly promising, as it suggests that fully automated pipelines can still achieve a reasonable level of CPS detection accuracy.
However, while CPS classification performance remains high with automatic segmentation, we found the number of instances decreased, as shown in Table \ref{tab:class-instances}. This is due to multiple oracle-segmented utterances being merged into a single segment under Google's system. We observed 518 such instances where multiple oracle segments were combined, with 18 cases merging more than two oracle segments. The underlying distribution of labels reveals important limitations of automated segmentation.

\begin{table*}[h]
\centering
\caption{Number of utterances and CPS labels under Oracle vs. Google segmentation.}
\begin{tabular}{rrrrr}
\toprule
& \textbf{Utterances}& \textbf{Const} & \textbf{Neg} & \textbf{Maintain} \\ \midrule
\textbf{Oracle Segments} &2482     & 906 & 866 & 391 \\
\textbf{Google Segments}  &1824  & 664 & 642 & 328 \\
\hline
\end{tabular}
\label{tab:class-instances}
\end{table*}

Merging distinct utterances into a single segment leads to multi-label segments and loss of granularity as seen in Figure \ref{fig:segments}, where the distinction between the two CPS facets is blurred. The merging of segments may obscure important micro-level interactions, potentially limiting an AI-driven system's ability to make fine-grained distinctions in collaborative behaviors, and other downstream tasks\cite{2024.EDM-long-papers.14,venkatesha2025propositional,castillon2022multimodal,nath2024any}.  While automated segmentation allows for efficient large-scale CPS classification, it sacrifices interpretability and label precision.


Beyond technical feasibility, the ability to automatically detect CPS behaviors has meaningful implications for classroom practice. If integrated into a teacher-facing dashboard, such a system could offer real-time or post-hoc summaries of group interactions, helping educators assess students’ collaboration skills. However, for such insights to be actionable, teachers need to know not just what behavior occurred, but who said what and when. Segmentation choices directly affect this interpretability: over-segmentation can fragment meaningful behaviors, while under-segmentation can blur speaker turns and behavior distinctions. Understanding this trade-off is crucial for real-world deployment. While our study focuses on Google’s segmentation system, which tends to under-segment, it does not examine other ASR pipelines that might over-segment or exhibit different boundary behavior. Exploring a wider range of segmentation strategies could yield substantially different outcomes. 
A single dataset (WTD) \cite{khebour_ibrahim_2023_8384960} was chosen for its annotated CPS labels and controlled task structure. While this ensures consistency, it limits generalizability to other collaborative contexts (e.g., open-ended tasks or diverse age groups). 
Additionally, our evaluation maps oracle CPS labels onto automatically segmented transcripts. This decision simplifies comparison across segmentation methods but different mapping strategies could result in varying classification performance. 

\subsubsection{Conclusion}
Our study demonstrates that fully automated pipelines for CPS detection utilizing ASR-based transcription and segmentation can achieve promising levels of performance with only minimal degradation compared to oracle transcriptions. This indicates that automated CPS monitoring systems in classroom settings are feasible.  
However, our study also highlights the need for more precise segmentation methods or post-processing strategies are needed to preserve fine-grained interaction. This work highlights the potential and limitations of AI-driven approaches for modeling CPS. 


\section{Acknowledgements}
This material is based in part upon work supported by the National Science Foundation (NSF) under
subcontracts to Colorado State University on award DRL 2019805 (Institute
for Student-AI Teaming), and by Other Transaction award HR00112490377 from the U.S. Defense Advanced Research Projects Agency (DARPA) Friction for Accountability in Conversational Transactions
(FACT) program. Approved for public release, distribution unlimited. Views expressed herein do not
reflect the policy or position of the National Science Foundation, the Department of Defense, or the U.S.
Government. All errors are the responsibility of the authors

\bibliographystyle{splncs04}
\bibliography{aied}

\begin{thebibliography}{10}
\providecommand{\url}[1]{\texttt{#1}}
\providecommand{\urlprefix}{URL }
\providecommand{\doi}[1]{https://doi.org/#1}

\bibitem{bergstra2013hyperopt}
Bergstra, J., Yamins, D., Cox, D.: Making a science of model search: Hyperparameter optimization in hundreds of dimensions for vision architectures. In: Proceedings of the 30th International Conference on Machine Learning (ICML). pp. 115--123 (2013)

\bibitem{blanchard2015study}
Blanchard, N., Brady, M., Olney, A.M., Glaus, M., Sun, X., Nystrand, M., Samei, B., Kelly, S., D’Mello, S.: A study of automatic speech recognition in noisy classroom environments for automated dialog analysis. In: Artificial Intelligence in Education: 17th International Conference, AIED 2015, Madrid, Spain, June 22-26, 2015. Proceedings 17. pp. 23--33. Springer (2015)

\bibitem{bradford2022deep}
Bradford, M., Hansen, P., Ross, J.B., Krishnaswamy, N., Blanchard, N.: A deep dive into microphone hardware for recording collaborative group work. In: Educational Data Mining Conference. Zenodo (2022)

\bibitem{bradford_automatic_2023}
Bradford, M., Khebour, I., Blanchard, N., Krishnaswamy, N.: Automatic detection of collaborative states in small groups using multimodal features. In: International Conference on Artificial Intelligence in Education. pp. 767--773. Springer (2023)

\bibitem{breiman2001random}
Breiman, L.: Random forests. Machine Learning  \textbf{45}(1),  5--32 (2001). \doi{10.1023/A:1010933404324}

\bibitem{cao2023comparative}
Cao, J., Ganesh, A., Cai, J., Southwell, R., Perkoff, E.M., Regan, M., Kann, K., Martin, J.H., Palmer, M., D’Mello, S.: A comparative analysis of automatic speech recognition errors in small group classroom discourse. In: Proceedings of the 31st ACM Conference on User Modeling, Adaptation and Personalization (UMAP ’23). pp. 250--261. ACM, Limassol, Cyprus (2023). \doi{10.1145/3565472.3595606}

\bibitem{castillon2022multimodal}
Castillon, I., Venkatesha, V., VanderHoeven, H., Bradford, M., Krishnaswamy, N., Blanchard, N.: Multimodal features for group dynamic-aware agents. In: Interdisciplinary Approaches to Getting AI Experts and Education Stakeholders Talking Workshop at AIEd. International AIEd Society (2022)

\bibitem{devlin2018bert}
Devlin, J.: Bert: Pre-training of deep bidirectional transformers for language understanding. arXiv preprint arXiv:1810.04805  (2018)

\bibitem{efron1993bootstrap}
Efron, B., Tibshirani, R.J.: An Introduction to the Bootstrap. Springer (1993)

\bibitem{evanini2013asr}
Evanini, K., Higgins, D., Zechner, K.: Automated speech scoring for non-native middle school students with multiple task types. In: Proceedings of Interspeech (2013)

\bibitem{openFeatures}
Eyben, F., Scherer, K.R., Schuller, B.W., Sundberg, J., André, E., Busso, C., Devillers, L.Y., Epps, J., Laukka, P., Narayanan, S.S., Truong, K.P.: The geneva minimalistic acoustic parameter set (gemaps) for voice research and affective computing. IEEE Transactions on Affective Computing  \textbf{7}(2),  190--202 (2016). \doi{10.1109/TAFFC.2015.2457417}

\bibitem{eyben2010opensmile}
Eyben, F., W{\"o}llmer, M., Schuller, B.: Opensmile: the munich versatile and fast open-source audio feature extractor. In: Proceedings of the 18th ACM international conference on Multimedia. pp. 1459--1462 (2010)

\bibitem{fawcett2006introduction}
Fawcett, T.: An introduction to roc analysis. Pattern Recognition Letters  \textbf{27}(8),  861--874 (2006)

\bibitem{Flor2016}
Flor, M., Yoon, S.Y., Liu, O.L., Wagner, M.: Automated classification of collaborative problem solving interactions in simulated science tasks. ETS Research Report Series  \textbf{2016}(1),  1--12 (2016)

\bibitem{freund1997decision}
Freund, Y., Schapire, R.E.: A decision-theoretic generalization of on-line learning and an application to boosting. Journal of Computer and System Sciences  \textbf{55}(1),  119--139 (1997). \doi{10.1006/jcss.1997.1504}

\bibitem{gales2008hmm}
Gales, M.J., Young, S.J.: The application of hidden markov models in speech recognition. Foundations and Trends in Signal Processing  \textbf{1}(3),  195--304 (2008)

\bibitem{jensen2023collaborative}
Jensen, A.K., et~al.: Collaborative problem solving: A pedagogy for workplace relevance. Nordic Journal of Vocational Education and Training  \textbf{13}(2),  45--73 (2023)

\bibitem{khebour_ibrahim_2023_8384960}
Khebour, I., Brutti, R., Dey, I., Dickler, R., Sikes, K., Lai, K., Bradford, M., Cates, B., Hansen, P., Jung, C., Wisniewski, B., Terpstra, C., Hirshfield, L., Puntambekar, S., Blanchard, N., Pustejovsky, J., Krishnaswamy, N.: {The Weights Task Dataset: A Multimodal Dataset of Collaboration in a Situated Task} (Sep 2023). \doi{10.5281/zenodo.8384960}, \url{https://doi.org/10.5281/zenodo.8384960}

\bibitem{khebour2024text}
Khebour, I., Brutti, R., Dey, I., Dickler, R., Sikes, K., Lai, K., Bradford, M., Cates, B., Hansen, P., Jung, C., et~al.: When text and speech are not enough: A multimodal dataset of collaboration in a situated task. Journal of open humanities data  \textbf{10}(1) (2024)

\bibitem{thinkkids}
Kids, T.: Collaborative problem solving. https://thinkkids.org/Schools/  (2025)

\bibitem{klieme2016assessing}
Klieme, E., Hartig, J., Rauch, D., Blum, W.: Assessing collaborative problem solving: An overview of the pisa 2015 assessment framework. In: Collaborative problem solving: An educational perspective. pp. 31--53. Springer (2016)

\bibitem{manuvinakurike2015asr}
Manuvinakurike, R., DeVault, D.: Using asr word confusion networks for modeling decisions in spoken dialogue systems. In: Proceedings of Interspeech (2015)

\bibitem{nath2024any}
Nath, A., Venkatesha, V., Bradford, M., Chelle, A., Youngren, A., Mabrey, C., Blanchard, N., Krishnaswamy, N.: Any other thoughts, hedgehog? linking deliberation chains in collaborative dialogues. arXiv preprint arXiv:2410.19301  (2024)

\bibitem{oecd2017pisa}
OECD: Pisa 2015 assessment and analytical framework: Science, reading, mathematic and financial literacy. Paris: OECD Publishing  (2017)

\bibitem{sun_towards_2020}
Sun, C., Shute, V.J., Stewart, A., Yonehiro, J., Duran, N., D'Mello, S.: Towards a generalized competency model of collaborative problem solving. Computers \& Education  \textbf{143},  103672 (Jan 2020). \doi{10.1016/j.compedu.2019.103672}, \url{https://www.sciencedirect.com/science/article/pii/S0360131519302258}

\bibitem{2024.EDM-long-papers.14}
Venkatesha, V., Nath, A., Khebour, I., Chelle, A., Bradford, M., Tu, J., Pustejovsky, J., Blanchard, N., Krishnaswamy, N.: Propositional extraction from natural speech in small group collaborative tasks. In: PaaÃŸen, B., Epp, C.D. (eds.) Proceedings of the 17th International Conference on Educational Data Mining. pp. 169--180. International Educational Data Mining Society, Atlanta, Georgia, USA (July 2024). \doi{10.5281/zenodo.12729792}

\bibitem{venkatesha2025propositional}
Venkatesha, V., Nath, A., Khebour, I., Chelle, A., Bradford, M., Tu, J., VanderHoeven, H., Bhalla, B., Youngren, A., Pustejovsky, J., et~al.: Propositional extraction from collaborative naturalistic dialogues. Journal of educational data mining  \textbf{17}(1),  183--216 (2025)

\bibitem{yoon2020asr}
Yoon, S.Y., Xue, Y., Warschauer, M.: Speech-to-text for literacy: How asr errors affect educational applications. In: Proceedings of the 58th Annual Meeting of the Association for Computational Linguistics (ACL). pp. 5059--5070 (2020)

\bibitem{yu2014asr}
Yu, D., Deng, L.: Automatic Speech Recognition: A Deep Learning Approach. Springer (2014)

\bibitem{zechner2015asr}
Zechner, K., Evanini, K., Yoon, S.Y., Wang, X.: The challenges of asr in automated speaking assessment. In: Proceedings of the Workshop on Speech and Language Technology in Education (SLaTE) (2015)

\end{thebibliography}

\end{document}